\documentclass{svproc}

\usepackage{url}

\usepackage{graphicx}
\usepackage{amsmath}
\usepackage{siunitx}
\usepackage[sicmds,freestanding]{hepunits}

\usepackage[T1]{fontenc}
\usepackage[utf8]{inputenc}
\usepackage{lmodern} 

\usepackage[numbers]{natbib}
\usepackage{subcaption}
\usepackage{caption}
\usepackage[capitalize]{cleveref}

\newcommand{\eV}{\ensuremath{\text{e\kern-0.1em V}}}
\newcommand{\EeV}{\ensuremath{\text{E\kern-0.01em e\kern-0.1em V}}}

\DeclareSIUnit{\year}{yr}

\title{The Pierre Auger Observatory: Results and Prospects}
\titlerunning{The Pierre Auger Observatory}  

\author{Qader Dorosti Hasankiadeh\inst{1}, for the Pierre Auger Collaboration\inst{2}}
\authorrunning{Qader Dorosti Hasankiadeh, for the Pierre Auger Collaboration}  

\institute{
    Center for Particle Physics Siegen, Department für Physik, Universität Siegen, Walter-Flex-Str. 3, 57072 Siegen, Germany 
    \and Full author list: \url{https://www.auger.org/archive/authors_2024_10.html} \\
    \email{dorosti@hep.physik.uni-siegen.de}
}

\begin{document}
\mainmatter              

\maketitle              

\begin{abstract}
The Pierre Auger Observatory advances the study of ultra-high-energy cosmic rays through a hybrid system of surface and fluo\-rescence detectors. This paper presents recent results, including refined spectrum measurements, anisotropy evidence, and new insights into cosmic-ray composition. Studies at energies beyond terrestrial accelerators reveal implications for particle physics. The AugerPrime upgrade will further enhance particle identification and extend the sensitivity to photons and neutrinos, broadening the Observatory’s capability to explore cosmic-ray sources and propagation, paving the way for new discoveries.

\end{abstract}

\section{Introduction}
The Pierre Auger Observatory~\cite{auger} in Mendoza, Argentina, is the world’s largest facility for studying ultra-high-energy cosmic rays (UHECRs). With a 3,000 km$^2$ detection area, it enables detailed studies of cosmic-ray origins, composition, and propagation in the southern hemisphere, supported by over 400 researchers from 95 institutions across 18 countries.

\begin{figure}[ht] 
    \centering
    \includegraphics[width=0.6\textwidth]{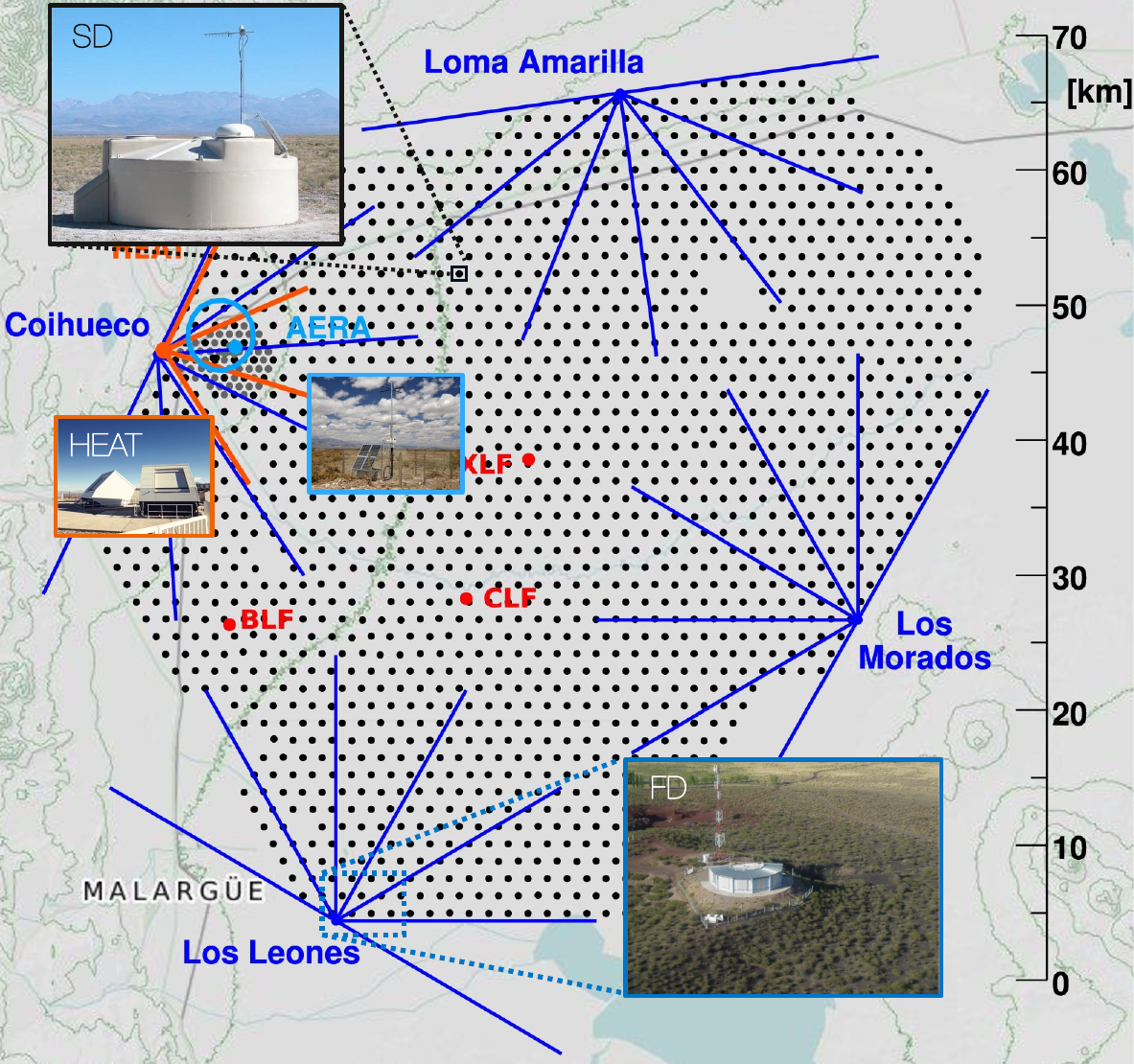} 
    \caption{Layout of the Pierre Auger Observatory, showing the positions of SD stations (black dots), the azimuthal fields of view of the FD (blue and red lines), with insets of selected detector images. This figure is adapted from~\cite{auger}.}
    \label{fig:augermap} 
\end{figure}

The main detection systems of the Auger Observatory include Surface Detector (SD), an array of 1660 stations, and the Fluorescence Detector (FD), organised into four installations around its perimeter (see~\cref{fig:augermap}). The SD stations, predominantly on a 1.5 km grid, detect ground-level particles from air showers, while the FD monitors ultraviolet emissions from nitrogen excited by these showers. The Infill region, with a denser grid of SD stations, integrates Underground Muon Detector (UMD), High Elevation Auger Telescopes (HEAT), and the Auger Engineering Radio Array (AERA), enhancing sensitivity to primary particle properties by providing detailed measurements of extensive air showers at lower energies. SD collects continuous ground-level data, while FD, operating 15\% of the time, captures shower profiles intermittently. SD-based energy estimates are influenced by model uncertainties, but calibrating with FD data reduces these effects, improving accuracy and understanding of UHECRs~\cite{calib1,calib2}. 

This paper highlights recent findings, the long-standing “muon puzzle,” and AugerPrime upgrades that promise further advancements in particle identification and cosmic-ray physics.

\section{Energy Spectrum}
\begin{figure}[ht]
    \centering
    \begin{subfigure}[t]{0.45\textwidth}
        \includegraphics[width=\linewidth]{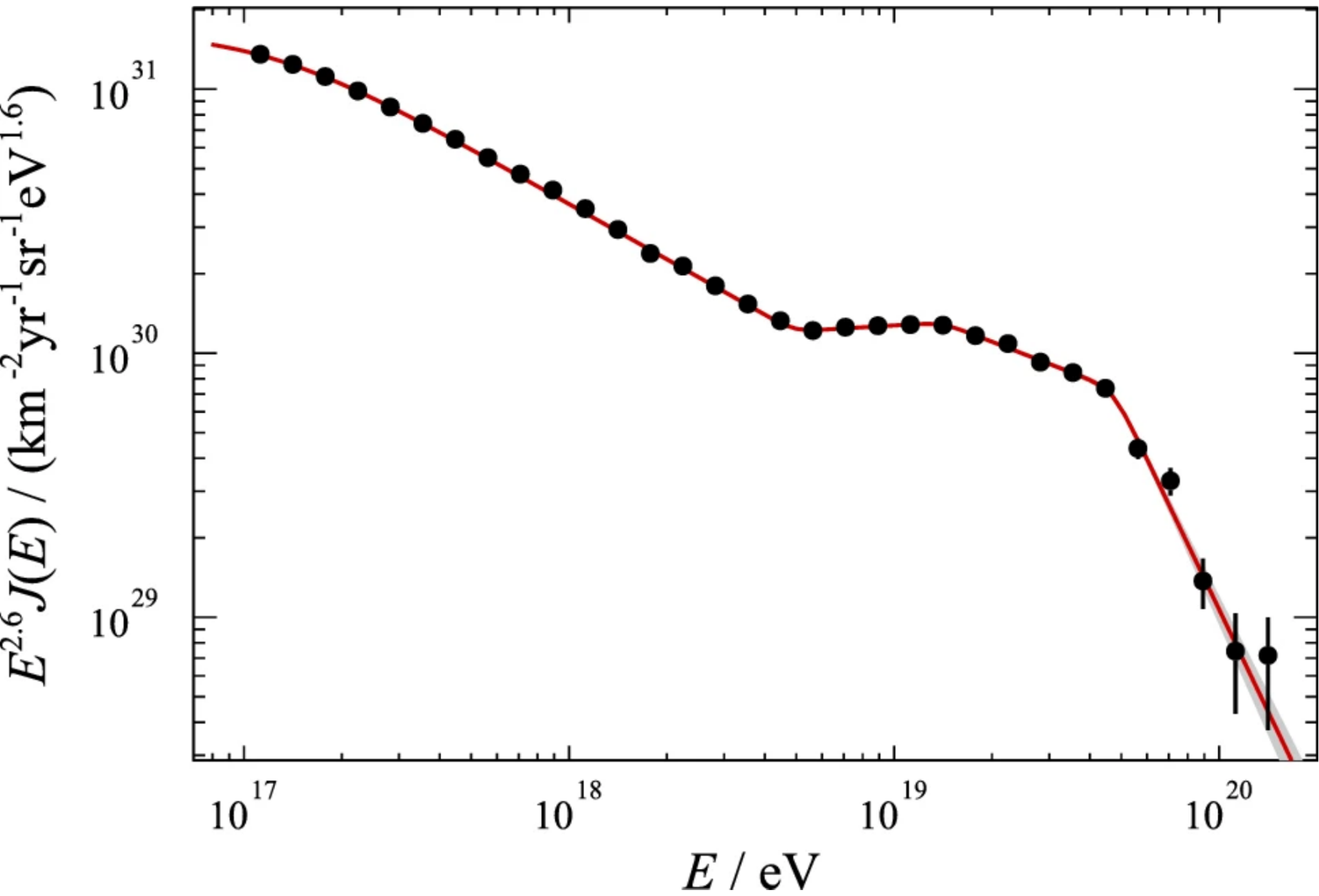}
        \caption{Energy spectrum with red overlay showing a fit with rescaling factors and uncertainties in grey~\cite{Abreu_2021}.}
        \label{fig:spectrum}
    \end{subfigure}
    \hfill
    \begin{subfigure}[t]{0.48\textwidth}
        \includegraphics[width=\linewidth]{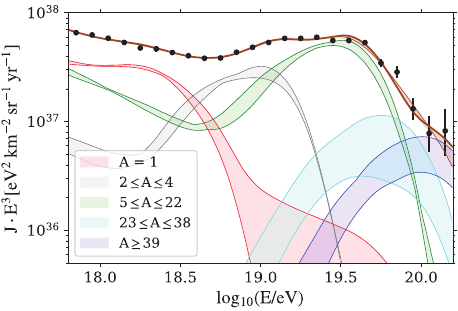}
        \caption{Combined mass fit of the all-particle energy spectrum, showing uncertainties from hadronic interaction models~\cite{compspectrum}.}
        \label{fig:compos_spectrum} 
    \end{subfigure}
    \caption{Energy and mass composition spectra for UHECRs measured by Auger.}
    \label{fig:combined}
\end{figure}

The UHECR energy spectrum measured by Auger (see~\cref{fig:spectrum}) reveals key insights into cosmic ray origins and propagation through hybrid measurements, without relying on simulations. This spectrum highlights distinct features that reflect the underlying acceleration mechanisms and source properties of UHECRs. 

Around \(10^{17}\, \eV\), a steepening known as the ``second knee'' suggests a reduction in contributions from galactic sources, particularly concerning heavy nuclei. At approximately \(10^{18.7}\, \eV\), the spectrum flattens at a point called the ``ankle'', likely marking a transition from cosmic rays of galactic origin to those of extragalactic origin. 

A newly observed feature around \( 10^{19.1}\, \eV \), referred to as the ``instep'', involves contributions from light and intermediate nuclei, potentially reflecting a shift in the population of extragalactic sources contributing to the spectrum. Beyond \(10^{19.7}\, \eV\), a significant suppression in the flux is observed. This suppression could be due to the Greisen–Zatsepin–Kuzmin (GZK) effect~\cite{greisen1966,zatsepin1966}, where UHECRs interact with cosmic microwave background photons, or it may indicate a natural limit to source acceleration. 

\section{Mass Composition}
The Observatory measures cosmic-ray composition by combining data from SD and FD. A key parameter in these analyses is the air shower maximum depth, \( X_{\text{max}} \), which correlates with the mass of the primary cosmic ray. Light nuclei, such as protons, tend to produce showers with deeper \( X_{\text{max}} \) values, while heavier nuclei, like iron, reach their maximum at shallower depths. By analysing \( X_{\text{max}} \) across different energy bins, Auger collaboration statistically determines the mass composition of cosmic rays at various energies.

\begin{figure}[ht] 
    \centering
    \includegraphics[width=0.85\textwidth]{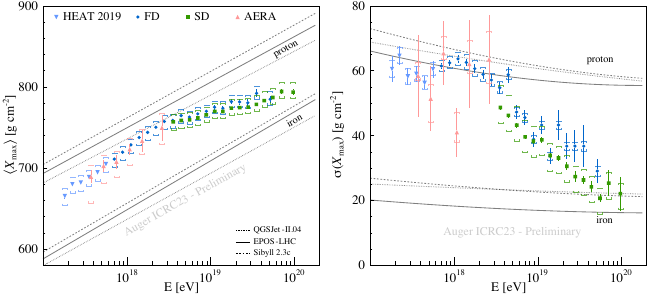} 
    \caption{First (left) and second (right) moments of $X_{max}$ distributions measured by the FD, SD, AERA, and HEAT (see~\cite{xmax} for more details).}
    \label{fig:xmax} 
\end{figure}

Observational data (~\cref{fig:xmax}) reveal a shift from lighter to heavier elements as energy increases. Specifically, below \SI{2}{\exa\eV}, the composition is dominated by lighter elements, which suggests a galactic origin. Above this energy threshold, however, the composition shifts towards heavier nuclei, indicative of extragalactic sources, which are thought to contribute more significantly to the high-energy cosmic-ray flux~\cite{xmax}. The second moment (variance) of \( X_{\text{max}} \) provides additional insights into the composition spread. At the highest energies, the second moment decreases, suggesting that the composition is becoming purer, possibly dominated by a narrower range of heavier nuclei. This narrowing of composition could reflect the nature of the extragalactic sources or the propagation effects experienced by cosmic rays as they travel across intergalactic space~\cite{xmax}.

A combined fit of the energy spectrum and mass composition data~\cite{compspectrum} reveals that cosmic rays below the ``ankle'' are predominantly composed of protons and intermediate-mass nuclei, while above this region, heavier nuclei become more prominent (see~\cref{fig:compos_spectrum}). This pattern supports the hypothesis of multiple extragalactic sources contributing to the highest energy cosmic rays. Such combined analyses, enhanced by the AugerPrime upgrade, will further reduce uncertainties and improve source differentiation.

\section{Muon Production Studies and the Muon Puzzle}
\begin{figure}[ht]
    \centering
    \includegraphics[width=0.6\textwidth]{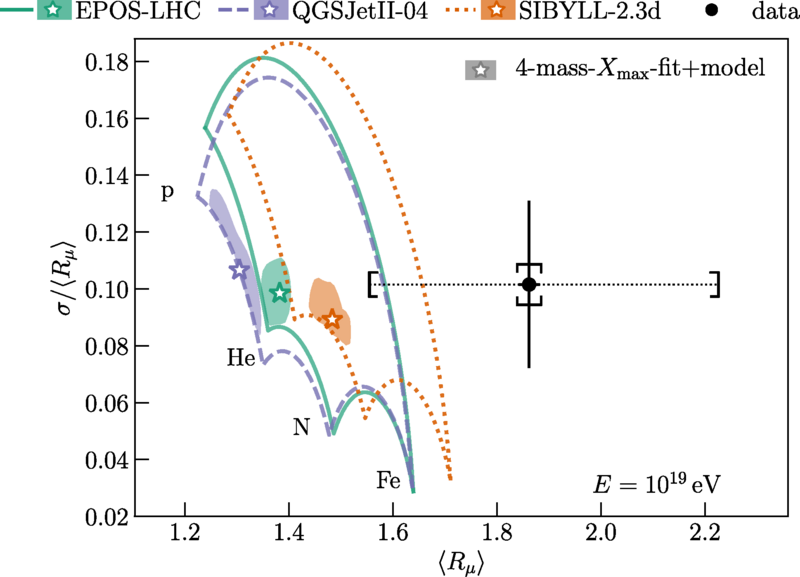}
    \caption{Muon fluctuations vs. average muon counts shown with statistical error bars and systematic brackets. Model predictions are in colour, \(X_{\text{max}}\)-derived predictions are marked with stars. The shaded areas show the uncertainty ranges~\cite{muon}.}
    \label{fig:muon}
\end{figure}

Data on inclined air showers from Auger provides valuable insight into muon production, as the electromagnetic component is absorbed at large zenith angles due to the extended interaction length. This allows for precise study of the muonic component, revealing an intriguing inconsistency known as the ``muon puzzle'': while the observed fluctuations in muon counts agree with model predictions, the average muon counts are consistently higher than expected (as shown in~\cref{fig:muon}).

This pattern—agreement in fluctuations but not in the mean count—suggests that the variability of muon production is accurately modeled, but there may be an underestimation in overall muon production within existing hadronic models at ultra-high energies. Given that these energies exceed those accessible by man-made accelerators, the Pierre Auger Observatory serves as a leading experiment in probing particle interactions at these extreme scales.

The ongoing AugerPrime upgrade, with its enhanced particle identification capabilities, will allow for more refined measurements of muon production. This will help to further disentangle model limitations and improve our understanding of particle interactions at the highest energies observed in nature, enhancing Auger’s contribution to high-energy astrophysics.

\section{Arrival Directions of Ultra-High-Energy Cosmic Rays}

The Pierre Auger Observatory has extensively analysed UHECR arrival directions to detect anisotropies and identify potential astrophysical sources at various angular scales. These studies leverage 19 years of data, with an accumulated exposure of about \SI[group-separator={,}]{135000}{\kilo\meter\squared\year\steradian}, offering a detailed view of the cosmic ray sky across a broad energy range. The analysis examines two angular scales of anisotropy:  intermediate for the highest energies and a large for lower energies.

\begin{figure}[ht]
    \centering
    \begin{subfigure}[t]{0.53\textwidth}
        \includegraphics[width=\linewidth]{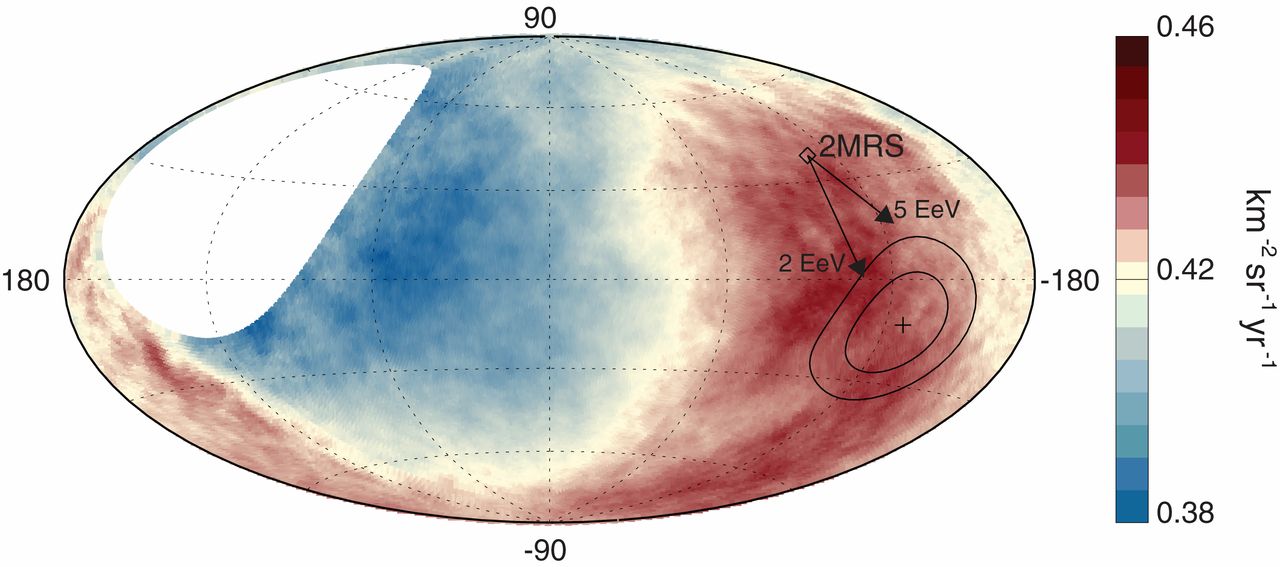}
        \caption{Cosmic-ray flux \( E \geq \SI{8}{\exa\eV} \) in Galactic coordinates, smoothed with a 45° top-hat, showing the dipole maximum (cross) with 68\%/95\% confidence contours; also displayed are the 2MRS galaxy dipole and magnetic field deflections for \( E/Z = 5 \) and \( \SI{2}{\exa\volt} \)~\cite{large_aniso}.}
        \label{fig:large_aniso}
    \end{subfigure}
    \hfill
    \begin{subfigure}[t]{0.4\textwidth}
        \includegraphics[width=\linewidth]{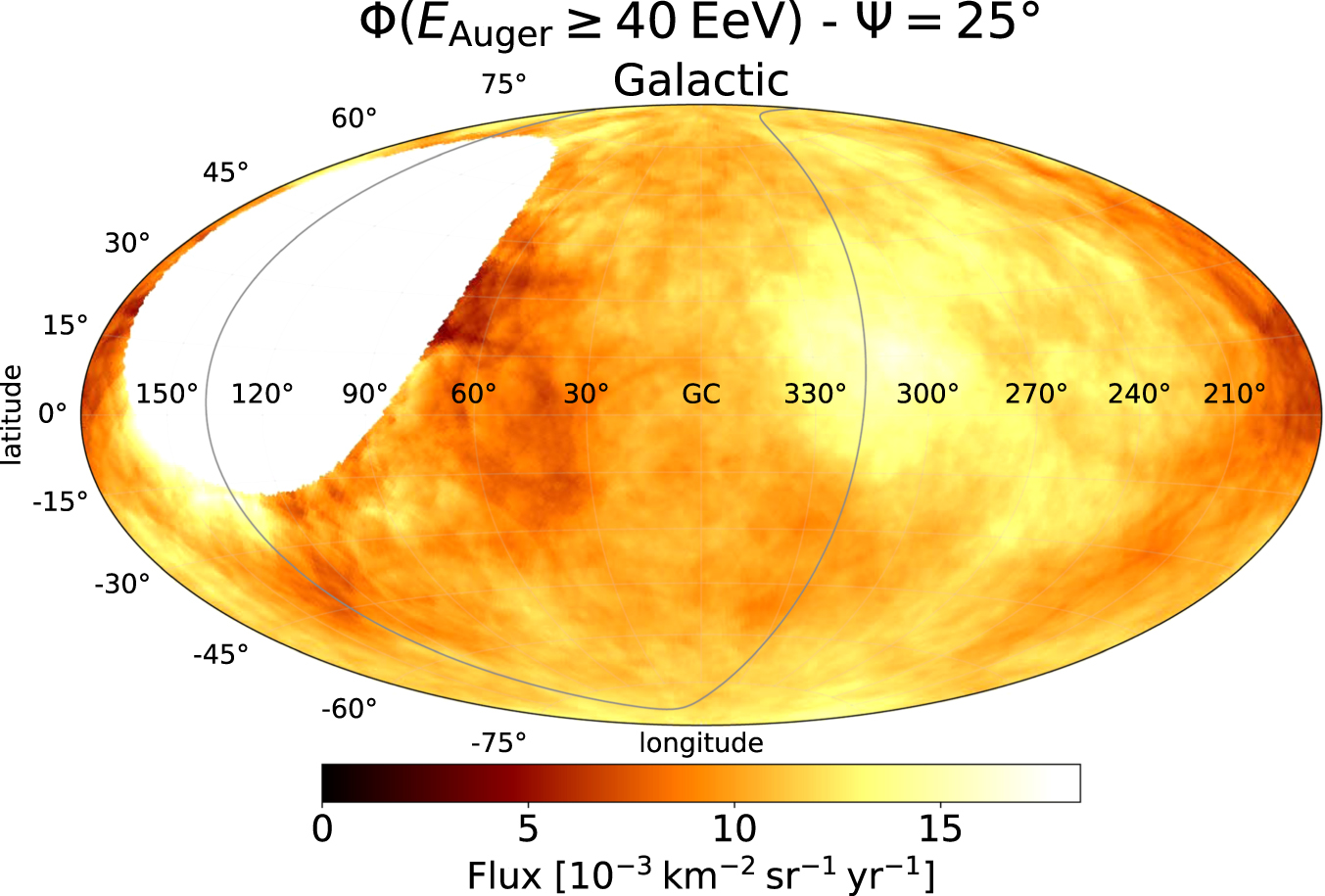}
        \caption{Sky map for energies $>\SI{40}{\exa\eV}$ with \SI{25}{\degree} top-hat smoothing. The supergalactic plane is shown with the grey line~\cite{inter_aniso}.}
        \label{fig:intermediate_aniso}
    \end{subfigure}
    \caption{Auger's sky maps for large-scale and intermediate-scale anisotropies.}
    \label{fig:combined2}
\end{figure}

A significant dipolar anisotropy above \SI{8}{\exa\electronvolt} is a key finding in \textit{large-scale searches}~\cite{large_aniso}. A Fourier analysis of right ascension and azimuth reveals a 3D dipole anisotropy with \(5.6\sigma\)  significance, with negligible higher-order moments, indicating a predominantly dipolar anisotropy. Galactic source models struggle to match these observations at high energies, as composition measurements do not indicate a dominant iron component. The flux-weighted 2MRS dipole points to \((\SI{251}{\degree}, \SI{38}{\degree})\), while the cosmic-ray flux direction differs by \SI{55}{\degree}. However, the galactic magnetic field is expected to shift direction and reduce dipole amplitude by \SI{70}{\percent}--\SI{90}{\percent} at \textit{rigidity} of 2--5\,\si{\exa\volt}, potentially bringing the directions into closer alignment (see the~\cref{fig:large_aniso}).

At \textit{intermediate scales} ($E > \SI{40}{\exa\electronvolt} $ see~\cref{fig:intermediate_aniso}), Auger data indicate an estimated \SI{10}{\percent} flux excess toward nearby multi-wavelength active galaxies~\cite{inter_aniso}. Analyses at a top-hat angular scale of $\psi =$ \SI{24}{\degree}--\SI{27}{\degree} optimised for clustering detection, reveal isotropy deviations of \(3.3\sigma\) and \(4.2\sigma\) for jetted AGN and starburst galaxy catalogues. An independent analysis of the Centaurus region, containing prominent active and star-forming galaxies, finds a \(4.1\sigma\) excess. With continued data accumulation at this scale, the exposure is expected to reach $\num[group-separator={,}]{165000} \pm \num[group-separator={,}]{15000}~\si{\kilo\meter\squared\year\steradian}$ by 2025, potentially raising the Centaurus excess significance to \(5\sigma\). However, the large angular radius limits conclusive links to Centaurus A or nearby starburst galaxies. An event-by-event analysis to distinguish light from heavy cosmic rays is needed; the ongoing AugerPrime upgrade aims to enable this, as discussed next.

\section{AugerPrime Upgrade, Future Prospects}
The nearing-completion AugerPrime upgrade~\cite{augerprime1,AugerPrime}, enables event-by-event particle identification. A scintillator-based surface detector installed atop each Water Cherenkov Detector (WCD) facilitates precise electron-to-muon ratio measurements, while a smaller photomultiplier tube in the WCD extends its dynamic range. The upgraded electronics board allows for improved data processing, and a new Radio Detector (RD) mounted on the WCD supports enhanced composition measurements, particularly for horizontal events. 
These advancements improve photon and neutrino detection, enabling AugerPrime to trace UHECR origins by distinguishing source from propagation effects, testing Lorentz invariance and physics beyond the Standard Model. Its enhanced capabilities promise groundbreaking insights into cosmic rays, potentially unveiling new physics and bringing us closer to unravelling their mysteries.


\end{document}